\documentclass{article}

\usepackage[final]{neurips_2021}

\usepackage[utf8]{inputenc} 
\usepackage[T1]{fontenc}    
\usepackage{hyperref}       
\usepackage{url}            
\usepackage{booktabs}       
\usepackage{amsfonts}       
\usepackage{nicefrac}       
\usepackage{microtype}      
\usepackage{xcolor}         
\usepackage{graphicx}
\usepackage{dirtytalk}
\usepackage{multirow}
\graphicspath{ {/Figures/} }

\title{AutoDC: Automated data-centric processing}

\author{Zac Yung-Chun Liu \thanks{Project website: \url{https://github.com/gohypergiant/research-AutoDC}}, Shoumik Roychowdhury, Scott Tarlow \\
\textbf{Akash Nair, Shweta Badhe, Tejas Shah} \\ \\
  Hypergiant Industries LLC \\
}
\begin{document}
\maketitle
\begin{abstract}
AutoML (automated machine learning) has been extensively developed in the past few years for the model-centric approach. As for the data-centric approach, the processes to improve the dataset, such as fixing incorrect labels, adding examples that represent edge cases, and applying data augmentation, are still very artisanal and expensive. Here we develop an automated data-centric tool (AutoDC), similar to the purpose of AutoML, aims to speed up the dataset improvement processes. In our preliminary tests on 3 open source image classification datasets, AutoDC is estimated to reduce roughly 80\% of the manual time for data improvement tasks, at the same time, improve the model accuracy by 10-15\% with the fixed ML code. 
\end{abstract}

\section{Introduction}

The availability of AutoML (automated machine learning) with publicly accessible pre-trained models enable domain experts to automatically build high-quality custom machine learning (ML) applications without much requirement for ML model construction knowledge [1, 2], which greatly speeds up the ML model development. AutoML has been an essential piece in the model-centric approach in the industry. As for the data-centric approach, the processes to improve the dataset, such as fixing incorrect labels, adding examples that represent edge cases, and applying data augmentation, are still very artisanal, and the automated tools for these tasks are lacking. Therefore, to facilitate these efforts, we develop an automated data-centric framework (AutoDC), similar to the purpose of AutoML, to enable domain experts to automatically and systematically improve datasets without much coding requirement and manual process. The remaining part of the paper is organized as follows. In Section 2, the ideation of this automated tool is described. In Section 3, the methodology is presented. In Section 4, the user workflow and AutoDC functionality are provided. In Section 5, the preliminary test results and their implications are discussed. The last section is the discussion for limitations, future works, and possible improvements.

\section{Ideation}

As Figure 1 shows, the common AutoML pipeline automates several processes, including data preparation and preprocessing, feature engineering, model generation, and hyperparameter tuning [3, 4]. Similarly, AutoDC is designed to automate the processes to improve dataset, including incorrect label correction, edge case detection, and augmentation, with a human-in-the-loop approach. By using the AutoML system, such as Google Cloud AutoML, domain experts only need to bring in the input data, and AutoML takes care of the manual ML processes, then produces output predictions, along with user-defined evaluation metrics (see Figure 1). With a similar idea, AutoDC is designed for domain experts to bring in a labeled dataset, such as annotated images, to the system; AutoDC takes care of the manual data improvement processes, and produces the improved dataset, by automatically correcting the incorrect labels (with user feedbacks), detecting edge cases, and augmenting edge cases. Note that AutoDC is still under development. Here we focus on the image datasets and image classification task as proof-of-concept (POC) in this paper. We plan to expand this methodology to other types of datasets, e.g. texts/ natural language process (NLP) as well as time-series.

\begin{figure}
  \includegraphics[scale=0.20]{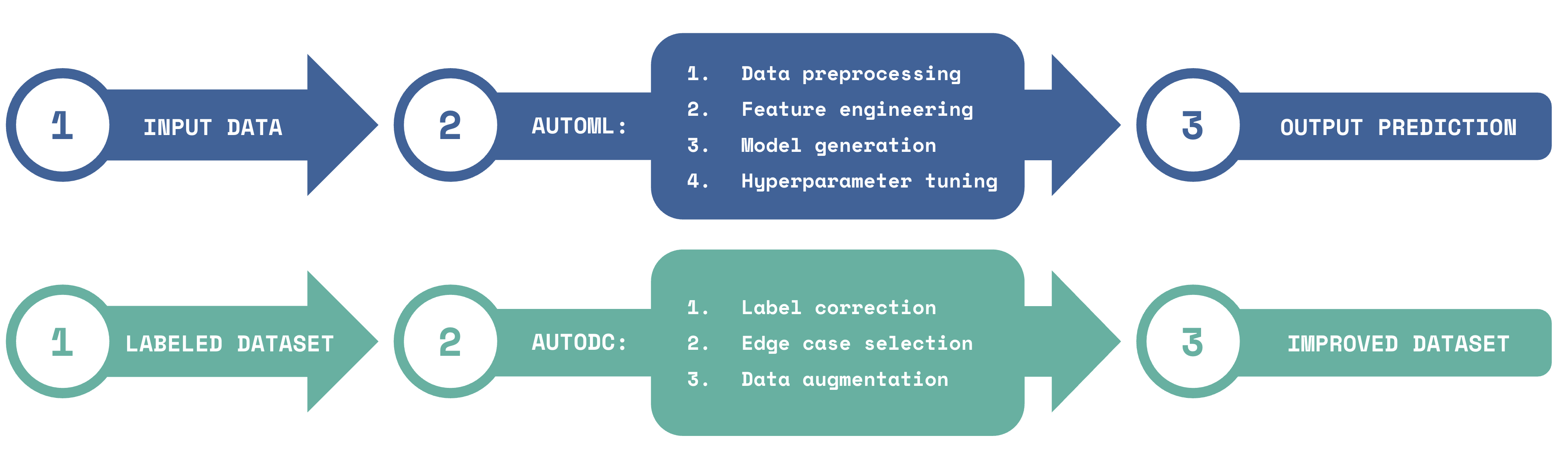}
  \centering
  \caption{AutoML functionality and the ideation of AutoDC tool.}
\end{figure}

\section{Methodology}
\subsection{Embeddings}

We utilize embeddings for the three dataset improvement routines: label correction, edge case detection, and data augmentation. An embedding is a relatively low-dimensional space translated from high-dimensional vectors [5, 6]. Embeddings represent the data in vector form that can be semantically grouped with their simility so they are useful for developing deep neural network models to tackle with large dimensional inputs; Word2Vec [7] and BERT [8] are a few famous examples. Here we generate image embeddings using pre-trained models, specifically ResNet50 [9] with ImageNet [10] pre-trained weights. We remove the last layer of sigmoid/softmax activation function of ResNet50 and pass image data to this last layer removed model to generate single-vector embeddings. We then apply dimensionality reduction technique, t-SNE (t-Distributed Stochastic Neighbor Embedding) [11, 12], to the embeddings so we can visualize the high-dimensional data in the form of clusters. Figure 2a and 2b show an example of embedding clusters for the roman numerals data “i” class.

\subsection{Outlier detection}
After the embeddings are created for each class in the dataset, we apply Isolation Forest [13] for outlier detection. Isolation Forest is an unsupervised model based on decision trees. In an Isolation Forest, randomly sub-sampled data is processed in a tree structure. With randomly selected features, the samples that end up in shorter branches are considered anomalies as it is easier for the tree to separate them from other observations. As for the samples that travel deeper into the tree are less likely to be anomalies as they require more cuts to isolate them. Figure 2c shows an example of outliers separated from the embedding clusters. These outliers are considered as either edge cases in the data or implied wrong labels. The red box in Figure 2c indicates the wrong label from the outliers.

\begin{figure}
  \includegraphics[scale=0.15]{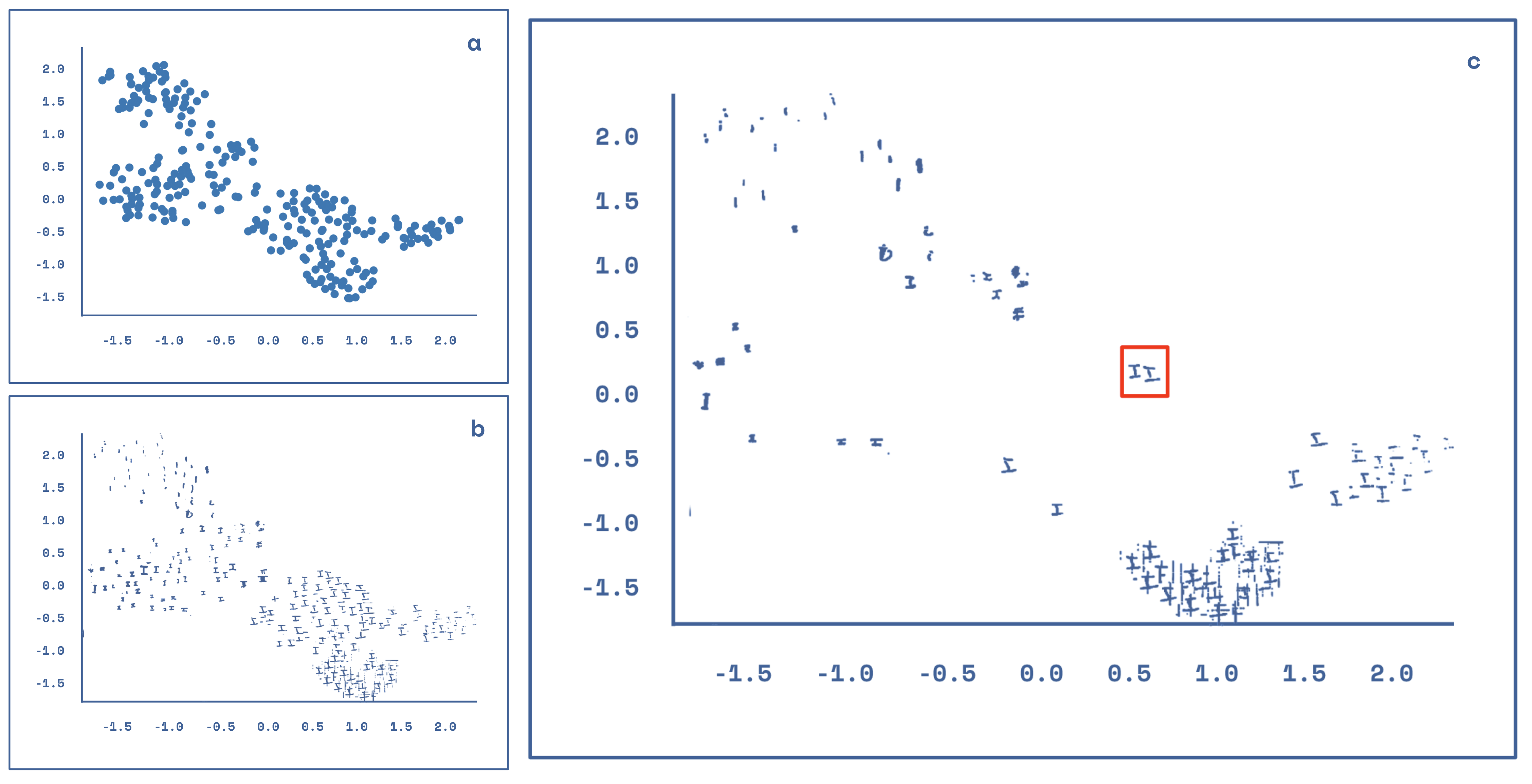}
  \centering
  \caption{(a) Image embedding of the roman numerals data “i” class in clustering view; blue dots represent each image data in class “i”. (b) The same embedding cluster as (a); the blue dots are replaced by actual images for visualization. (c) Edge case candidates recommended by Isolation Forest model. Red box indicates the wrong label as it should be labeled as class “ii”.}
\end{figure}

\section{Workflow and functionality}

AutoDC is designed to follow three dataset improvement steps/ routines sequentially: (1) label correction, (2) edge case selection, (3) data augmentation. Appendix Figure 1 summaries the workflow in AutoDC. As discussed in Section 2, before using AutoDC, we assume the users have identified a model (custom ML model or AutoML) that is well suited for the task at hand. AutoDC is designed for the users to gain additional ML performance by augmenting the dataset quality as a whole. Note that the data quality processes (e.g. embeddings creation, tSNE modeling, Isolation Forest modeling etc.) are automated, however, the downstream ML task is not, unless the users are using an AutoML framework for it.

\subsection{Label correction}

In AutoDC, users are required to supply labeled datasets, then the system automatically creates embeddings and detects outliers using the methods described in section 3. AutoDC then presents a visualization of image embeddings and outliers for each class. For the label correction function in AutoDC, we design it to request user feedback (i.e. human-in-the-loop). Users are prompted to examine the embedding outliers and correct the wrong labels (red box on Figure 2c).  

\subsection{Edge case Selection}
After correcting the wrong labels, users are prompted to select edge cases visually. The embedding outliers are the edge case candidates. This process can be subjective and user dependent. Therefore, we provide an option for the users to include all the edge case candidates recommended by Isolation Forest.  

\subsection{Data Augmentation}

After the edge case selection, users are prompted to select augmentation techniques that are proper for their datasets, including adding Gaussian noise, cropping, flipping, rotation, scaling, brightness, contrast, and saturation. We also provide an option for the users to include all augmentation techniques. The default value range for each augmentation technique is listed in Appendix Table 1. The augmentation here is mainly for the edge cases. The users are able to adjust the proportion of augmented edge cases for the output dataset.  

\section{Preliminary tests}
In this paper we tested three open source image classification datasets: (1) roman numerals [14], (2) Asirra (Animal Species Image Recognition for Restricting Access) for dogs and cats [15], and (3) Stanford parasitic snail for neglected tropical disease (NTD) [16, 17]. The details of these datasets are summarized in Table 1.

\begin{table}[h!]
\caption{Datasets used in this study.}
\centering
 \begin{tabular}{||c|c|c|c||} 
 \hline
\textbf{Dataset} & \textbf{Number of Images} & \textbf{Number of Classes} & \textbf{Ref}. \\ [0.5ex] 
 \hline\hline
 Roman numerals & $2,966$ & $4$ & $14$ \\[0.25ex] 
 \hline
  Asirra- dogs vs cats & $25,000$ & $2$ & $15$ \\[0.25ex] 
 \hline
  Stanford parasitic snail for NTD & $5,140$ & $4$ & $16,17$ \\[0.25ex] 
 \hline
 \end{tabular}

\label{table:1}
\end{table}

In the preliminary tests for this paper, we ran through all three dataset improvement routines in AutoDC with a single node CPU cloud instance. We found that less than 1 labels are incorrect in the datasets. As for edge cases, Isolation Forecast recommended 10-15\% of datasets to be edge case candidates and we included all the candidates in our tests. As for the proportion of augmented edge cases for the output datasets, we found that a 15-25\% ratio is optimal to get the best results from our image classification model. We implemented the ResNet50 model with pre-trained weights of ImageNet and ML code is fixed (train/test split: 80/20, learning rate=0.0001, batch size= 8, optimizer: Adam) in our preliminary tests. We first tested the model with the original/ unmodified labeled datasets (as control groups) and we then tested the improved datasets (as experiment groups) that came out from AutoDC, without fine-tuning the ResNet model. We found that the improved datasets boosted the model accuracy to 10-15\% in our test cases (Table 3). Since the dataset improvement tasks were done automatically with AutoDC, although it still requires some user feedback, we estimated the users saved 80\% of the time if all the improvement tasks were done manually, for example, going through every single image for wrong labels, manually examining and hand picking the edge cases in the datasets, and manually augmenting the images.

\subsection{Next Steps}
In our preliminary analysis, we only compared the classification model accuracy of the controlled/ unmodified datasets with fine-tuned datasets that went through all three dataset routine tasks (e.g. label correction, edge case selection, and augmentation). The obvious next step would be to examine the classification accuracy improvements with each fine-tuning routine that is done (1) independently of each other, (2) mutually exclusive combination of each other, and (3) total inclusion of all methods. This analysis would provide more insights on the efficiency of each routine and their correlation with different types of datasets. Another test case would be to take a corrupted dataset to go through AutoDC fine-tuning routines and compare the modeling results, which would further prove the usefulness of this automated data-centric approach.

\begin{table}[h!]
\caption{Preliminary test results (fixed ResNet50 model)}
\centering
\begin{tabular}{ |p{3.3cm}||p{2cm}|p{2cm}|p{2.2cm}|p{2cm}|} 
 \hline
Dataset & Wrong label ratio & Optimal aug ratio & Model acc- unmodified data & Model acc- improved data \\
 \hline\hline
 Roman numerals & $0.8\%$  & $25\%$ & $0.65$ & $0.80\pm0.05$\\
 \hline
 Asirra- dogs vs cats & $0.001\%$  & $20\%$ & $0.72$ & $0.82\pm0.03$\\
 \hline
 Stanford parasitic snail & $0.5\%$  & $15\%$ & $0.81$ & $0.95\pm0.01$\\
 \hline
\end{tabular}

\label{tabel:2}
\end{table}

\section{Limitations, improvement, and broader impact}
The limitations of the current AutoDC tool include (1) the users need to identify a training model or an AutoML framework in advance, (2) the process to fine tune the routine parameters, such as edge case selection ratio and augmentation ratio, is still manual, meaning how much of the improvement the model will get with the improved dataset came out from AutoDC cannot be known in advance.     Therefore, we identified several improvements in our user flow and methodology that can be made for AutoDC tools. First, an AutoML framework can be combined with AutoDC so the users are able to fine-tune the dataset routine parameters in the searching fashion to improve the dataset and the model concurrently and iteratively. Second, since most mature AutoML tools have integrated frontend user interface, we also aim to build such easy-to-use graphic user interface (GUI) to the dataset routines in AutoDC. Third, for the methodology improvement, the newer and more efficient computer vision models, such as EfficientNet [18], as well as other robust dimensionality reduction algorithms, such as UMAP (Uniform Manifold Approximation and Projection) [19], can be applied to create the image embeddings in AutoDC. In addition, more advanced augmentation approaches, such as GAN (Generative adversarial networks) [20], can be tested as well. Since AutoDC tools are built on top of open source ML applications and techniques, we can adopt the newly developed ML toolings along the way to ensure its efficiency and robustness for domain experts to prepare the improved quality of datasets so they can undertake the data-centric approach with ease.

\section*{Funding and acknowledgements}
The funding for this project was provided by Hypergiant Industries LLC. The authors thank Andy Armstrong, the Design Director at Hypergiant, for creating the plots and visualization.

\section*{References}

{
\small

[1] He, X., Zhao, K. \& Chu, X. (2021) AutoML: A Survey of the State-of-the-Art. Knowledge-Based Systems, 212, 106622.

[2] Zöller, M. A., \& Huber, M. F. (2021) Benchmark and survey of automated machine learning frameworks. Journal of Artificial Intelligence Research, 70, 409-472.

[3] Gijsbers, P., LeDell, E., Thomas, J., Poirier, S., Bischl, B., \& Vanschoren, J. (2019) An open source AutoML benchmark. arXiv preprint arXiv:1907.00909.

[4] Feurer, M., Eggensperger, K., Falkner, S., Lindauer, M., \& Hutter, F. (2018, July) Practical automated machine learning for the automl challenge 2018. In International Workshop on Automatic Machine Learning at ICML (pp. 1189-1232).

[5] Kiela, D., \& Bottou, L. (2014, October) Learning image embeddings using convolutional neural networks for improved multi-modal semantics. In Proceedings of the 2014 Conference on empirical methods in natural language processing (EMNLP) (pp. 36-45).

[6] Harsanyi, J. C., \& Chang, C. I. (1994) Hyperspectral image classification and dimensionality reduction: An orthogonal subspace projection approach. IEEE Transactions on geoscience and remote sensing, 32(4), 779-785.

[7] Mikolov, T., Chen, K., Corrado, G., \& Dean, J. (2013) Efficient estimation of word representations in vector space. arXiv preprint arXiv:1301.3781.

[8] Devlin, J., Chang, M. W., Lee, K., \& Toutanova, K. (2018) Bert: Pre-training of deep bidirectional transformers for language understanding. arXiv preprint arXiv:1810.04805.

[9] He, K., Zhang, X., Ren, S., \& Sun, J. (2016) Deep residual learning for image recognition. In Proceedings of the IEEE conference on computer vision and pattern recognition (pp. 770-778).

[10] Deng, J., Dong, W., Socher, R., Li, L. J., Li, K., \& Fei-Fei, L. (2009, June) Imagenet: A large-scale hierarchical image database. In 2009 IEEE conference on computer vision and pattern recognition (pp. 248-255). Ieee.

[11] Van der Maaten, L., \& Hinton, G. (2008) Visualizing data using t-SNE. Journal of machine learning research, 9(11).

[12] Van Der Maaten, L. (2014) Accelerating t-SNE using tree-based algorithms. The Journal of Machine Learning Research, 15(1), 3221-3245.

[13] Liu, F. T., Ting, K. M., \& Zhou, Z. H. (2008, December) Isolation forest. In 2008 eighth ieee international conference on data mining (pp. 413-422). IEEE.

[14] The roman numeral dataset. DOI: 10.5281/zenodo.5385144

[15] The Asirra dataset. \url{https://www.kaggle.com/c/dogs-vs-cats}

[16] Tallam, K., Liu, Z. Y. C., Chamberlin, A. J., Jones, I. J., Shome, P., Riveau, G., ... \& De Leo, G. A. (2021) Identification of Snails and Schistosoma of Medical Importance via Convolutional Neural Networks: A Proof-of-Concept Application for Human Schistosomiasis. Frontiers in Public Health, 900.

[17] Liu, Z. Y. C., Chamberlin, A. J., Shome, P., Jones, I. J., Riveau, G., Ndione, R. A., ... \& De Leo, G. A. (2019) Identification of snails and parasites of medical importance via convolutional neural network: an application for human schistosomiasis. bioRxiv, 713727.

[18] Tan, M., \& Le, Q. (2019, May) Efficientnet: Rethinking model scaling for convolutional neural networks. In International Conference on Machine Learning (pp. 6105-6114). PMLR.

[19] McInnes, L., Healy, J., \& Melville, J. (2018) Umap: Uniform manifold approximation and projection for dimension reduction. arXiv preprint arXiv:1802.03426.

[20] Goodfellow, I., Pouget-Abadie, J., Mirza, M., Xu, B., Warde-Farley, D., Ozair, S., ... \& Bengio, Y. (2014) Generative adversarial nets. Advances in neural information processing systems, 27.
}

\newpage
\section*{Appendix}
\setcounter{figure}{0} 
\setcounter{table}{0} 

\begin{figure}[!ht]
  \includegraphics[scale=0.17]{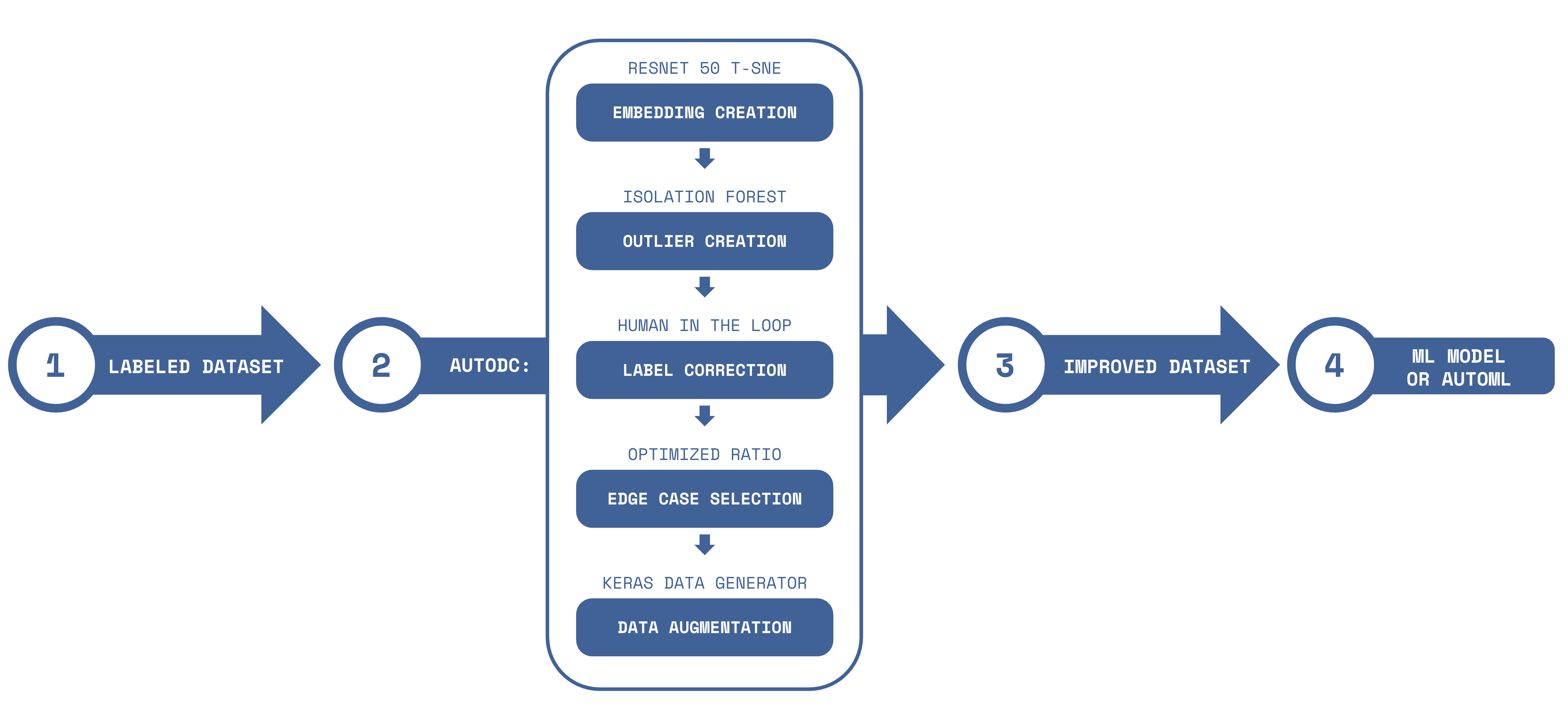}
  \centering
  \caption{AutoDC workflow.}
\end{figure}

\begin{table}[h!]
\caption{Default value range for the augmentation technique used in this study.}
\centering
 \begin{tabular}{||c|c||} 
 \hline
 \textbf{Augmentation Technique} & \textbf{Range} \\ [0.5ex] 
 \hline\hline
 Gaussian Noise Scale & $[10, 60]$ \\[0.25ex] 
 \hline
 Random Crop & $[0.0, 0.5]$ \\[0.25ex] 
 \hline
 Horizontal/ Vertical Flip & $90^{\circ}$\\[0.25ex] 
 \hline
 Rotation & $[30^{\circ}, 60^{\circ}]$ \\[0.25ex] 
 \hline
 Scaling & $[0.5, 1.0]$ \\[0.25ex] 
 \hline
 Brightness & $[0.2, 0.8]$\\[0.25ex] 
 \hline
 Contrast & $[0.1, 0.6]$\\[0.25ex] 
 \hline
 Saturation & $[0.1, 0.6]$\\[0.25ex] 

 \hline
 \end{tabular}
 
\label{table:1}
\end{table}

\end{document}